\documentclass[conference]{IEEEtran}
\IEEEoverridecommandlockouts

\usepackage{cite}
\usepackage{amsmath,amssymb,amsfonts}
\usepackage{algorithmic}
\usepackage{graphicx}
\usepackage{textcomp}
\usepackage{xcolor}
\usepackage{import}
\usepackage{cite}
\usepackage{booktabs}
\usepackage{comment}
\usepackage{enumerate}
\usepackage{array,booktabs,ragged2e}
\usepackage{multirow}
\usepackage{xcolor}
\usepackage{makecell}
\usepackage{authblk}

\begin{document}

\title{Counting Cards: Exploiting Variance and Data Distributions for Robust Compute In-Memory}

\author[1]{Brian Crafton$^\dagger$\thanks{\centering $\dagger$ These authors contributed equally}}
\author[1]{Samuel Spetalnick$^\dagger$}
\author[1]{Arijit Raychowdhury}
\affil[1]{Georgia Institute of Technology, Atlanta, GA}
\affil[1]{School of Electrical and Computer Engineering}
\affil[ ]{brian.crafton@gatech.edu, arijit.raychowdhury@ece.gatech.edu \vspace{-0.5cm}}

\maketitle

\begin{abstract}
Compute in-memory (CIM) is a promising technique that minimizes data transport, maximizes memory throughput, and performs computation on the bitline of memory sub-arrays. 
Utilizing embedded non-volatile memories (eNVM) such as resistive random access memory (RRAM), various forms of neural networks can be implemented.
Unfortunately, CIM faces new challenges traditional CMOS architectures have avoided.
In this work, we explore the impact of device variation and propose a new algorithm based on device variance to increase both performance and accuracy for CIM designs. 
We demonstrate a 36\% power improvement and 31\% performance improvement, while satisfying a programmable error threshold.
\end{abstract}

\section{Introduction} \label {section:intro}


Over the last decade, tremendous progress towards accelerating machine learning workloads has been made at all levels of the computing hierarchy, enabling orders of magnitude improvement in energy efficiency.
%
At the software level, models are compressed, pruned, and quantized to minimize the total size of the model and cost of a single inference \cite{hubara2017quantized}.
At the hardware level, prior work focuses on maximizing the reuse of all data such that expensive memory accesses and total data movement is minimized \cite{chen2017eyeriss}.
%
%
%
Both of these strategies focus on minimizing the cost of data movement and memory accesses, while maximizing the utility of available on-chip memory capacity. 
While these techniques yield strong results, they still face the fundamental technological limitations of CMOS. 
In particular, the large size of the SRAM bitcell ($\approx 100-150F^2$) results in limited on-die capacity, which necessitates movement of data from an external DRAM to the on-die SRAM at more energy per bit.

Fortunately a new class of high density eNVM is positioned to minimize data movement and increase memory throughput by performing CIM. 
CIM seeks to perform matrix multiplication ($\vec{y} = W \vec{x}$) in a crossbar structure in the analog domain using Ohm's law, exploiting the non-volatile conductance state(s) provided by the non-volatile memory.
Using this technique, each weight of the matrix ($W_{ij}$) is programmed as the conductance of a bit-cell and each value of the vector ($\vec{x_i}$) is converted to a corresponding voltage and applied to the rows of the memory crossbar. 
The current through each cell is proportional to the product of the programmed conductance ($W_{ij}$) and applied voltage ($\vec{x}_i$) (Ohm's Law). 
By Kirchhoff’s current law (KCL), the resulting currents that are summed along the columns of the crossbar are proportional to the product of the matrix and vector, ($\vec{y}$).
In this procedure, the only data transport required for matrix multiplication is the feature vector ($\vec{x}$) from memory and result ($\vec{y}$) to the memory. 
Therefore, CIM enables in-place computing, thereby eliminating the majority of the data transfer and energy cost of data intensive operations.

Despite these benefits, a key obstacle in designing a reliable CIM accelerator with eNVM is the inherent cell-to-cell variance in the device's resistive state. 
These variances are not specific to eNVM, and occur due to process and temperature or write-to-write (cycle-to-cycle) variations.
Conventional digital memory such as SRAM overcomes this challenge using differential sensing and a large ratio between the `0' and `1' states. 
However, when reading multiple memory cells at the same time with an analog-to-digital converter (ADC), high variance between resistive states results in sum-of-products errors accumulated on the bitline. 
Given that these operations are used to implement matrix multiplication, and thus neural networks, we find that device level variance results in erroneous computation. 
While neural networks can tolerate these errors to some extent, accuracy degrades as a function of the error rate. 

%
Recent work has attempted to mitigate the impact of these errors in several different ways. 
Training a network to be robust to variance induced error can be done both off-chip and on-chip. 
Off-chip training attempts to train a neural network to tolerate variance induced errors \cite{long2019design}, however this technique still results in accuracy degradation. 
On-chip training \cite{sun2019impact} can be done to minimize error for a specific chip, however this is expensive and will still ultimately result in accuracy degradation. 
Expensive write-verify methods have been proposed to reduce the cell-to-cell variance \cite{wu201840nm}. 
However these methods require high write energy and variable latency, and greatly reduce the endurance of the device.

In this work, we identify techniques that can be used at the circuit level to mitigate the cell-to-cell variance of eNVM. 
We quantify the relationship between cell-to-cell variance and the ADC precision and their role on the accuracy and performance of CIM operations.
Like prior work, we find that there exists a trade-off between accuracy and performance \cite{rekhi2019analog}. 
Building upon this, we use statistical estimation based on the variance and weight distributions to find optimal throughput for CIM arrays under an error constraint. 
For our experiments we simulate the performance, power, and accuracy of a CIM accelerator on a typical convolutional neural network (CNN) workload.
Although we apply our techniques to deep learning, we claim that the techniques we propose can be extended to any CIM application that suffers from variance.
Our results reveal power and performance improvements over commonly used compute-in memory techniques. 
We demonstrate a 36\% power improvement and 31\% performance improvement, while satisfying a programmable error threshold.


\section{Background and Motivation} \label{section:background}

\begin{figure}
\centering
\includegraphics[width=0.48\textwidth]{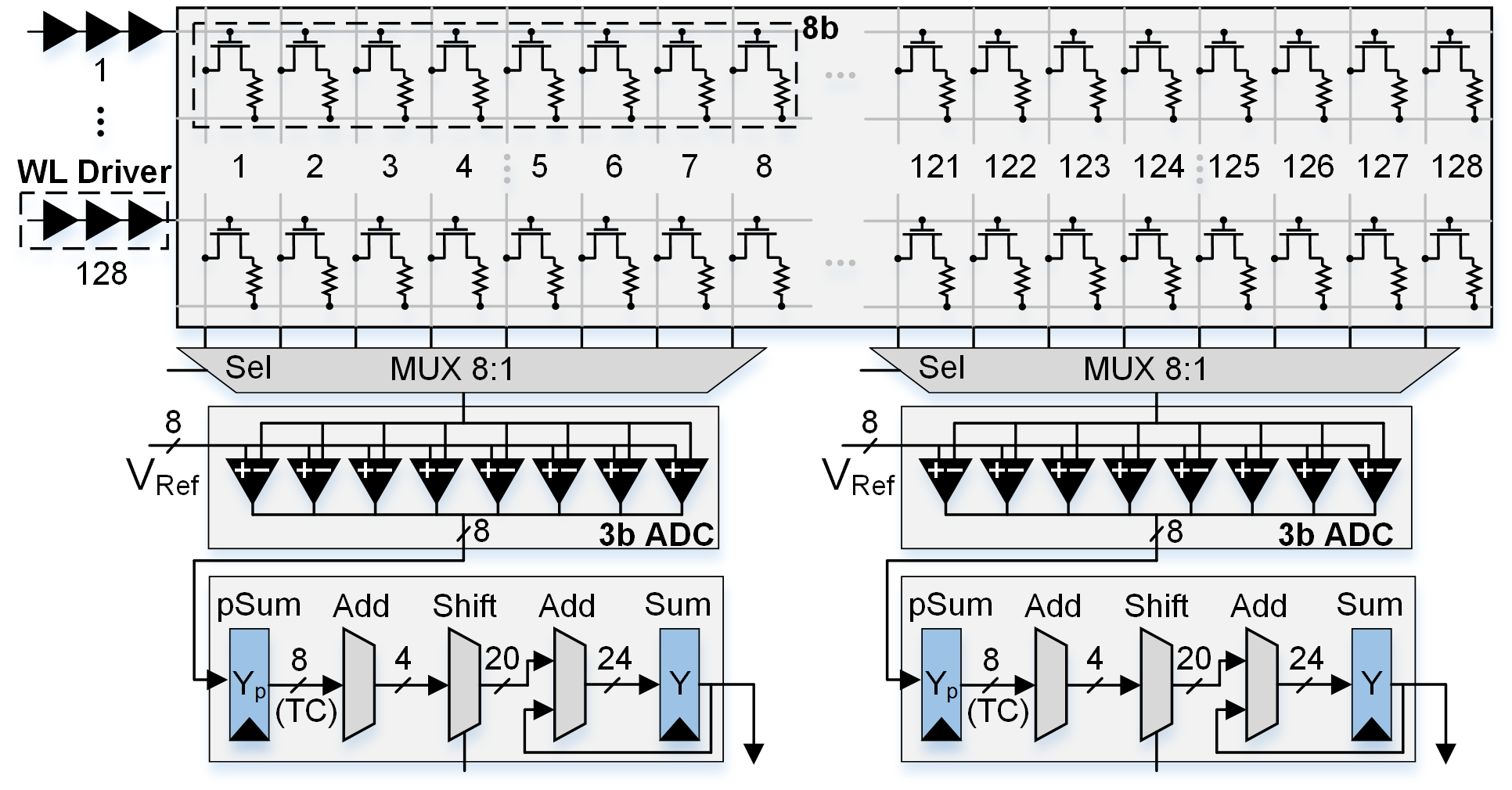}
\vspace{-0.25cm}
\caption{
128$\times$128 RRAM array architecture. 8 adjacent cells form an 8b weight and share a 3b ADC through a 8-to-1 multiplexer. Shift and add logic accumulate partial sums and apply corresponding magnitude.}
\label{fig:array}
\end{figure}

CIM systems use binary or multi-level cells as weights to perform matrix multiplication in memory \cite{shafiee2016isaac}. 
Assuming 8b matrix multiplication and binary cells, we must use 8 adjacent cells to form a single 8-bit weight, like those shown in the columns of Figure~\ref{fig:array}. 
The 8-bit vector inputs to this array are represented as voltages (GND and VDD), and input one at a time over 8 cycles.
The binary dot product is the current passing through each cell, and the sum-of-products is accumulated along the bitline. 
The resulting binary dot product is then collected at the ADCs.
If there are more inputs then states the ADC can distinguish, we must divide the binary dot product over multiple cycles and sum the partial products together with CMOS.
For each input bit and weight bit binary product, we perform this binary dot product, and then multiply (shift) the result by the sum of the magnitudes of each bit.  
Thus in order to perform an 8b multiplication, we perform binary multiplication between each bit in the input data and each bit in the weight data, and then shift by combined magnitude.
In this way, we can perform 8-bit matrix multiplication by performing 64 binary multiplications with shift and add operations.

\subsection{Compute In-Memory (CIM)}

There are two common techniques for performing compute in memory. 
The first technique, we call \textit{baseline}, is simply reading as many rows as the ADC precision allows (e.g. for a 3-bit ADC, we read 8 rows simultaneously). 
The next technique is commonly called zero skipping \cite{peng2019optimizing}, where only rows with `1's are read. 
This technique exploits sparsity in the input features or activations (for neural networks). 
Zero skipping performs faster than the baseline technique because for most cases it will process more total rows per cycle.
By encapsulating the array, ADCs, and shift and add logic, a matrix multiplication engine can be created. 
For example, if a 128$\times$128 array is used, a 128$\times$16 8b matrix multiplication can be implemented since 16 8b words can be fit across the 128 bitlines.
Using these arrays as building blocks, prior work has implemented Convolutional Neural Networks (CNN) \cite{shafiee2016isaac} and Recurrent Neural Networks (RNN) \cite{long2019design} where a group of arrays implement a larger matrix multiplication. 

\subsection{Quantifying the Impact of Variance}

\begin{figure}
\centering
\includegraphics[width=0.48\textwidth]{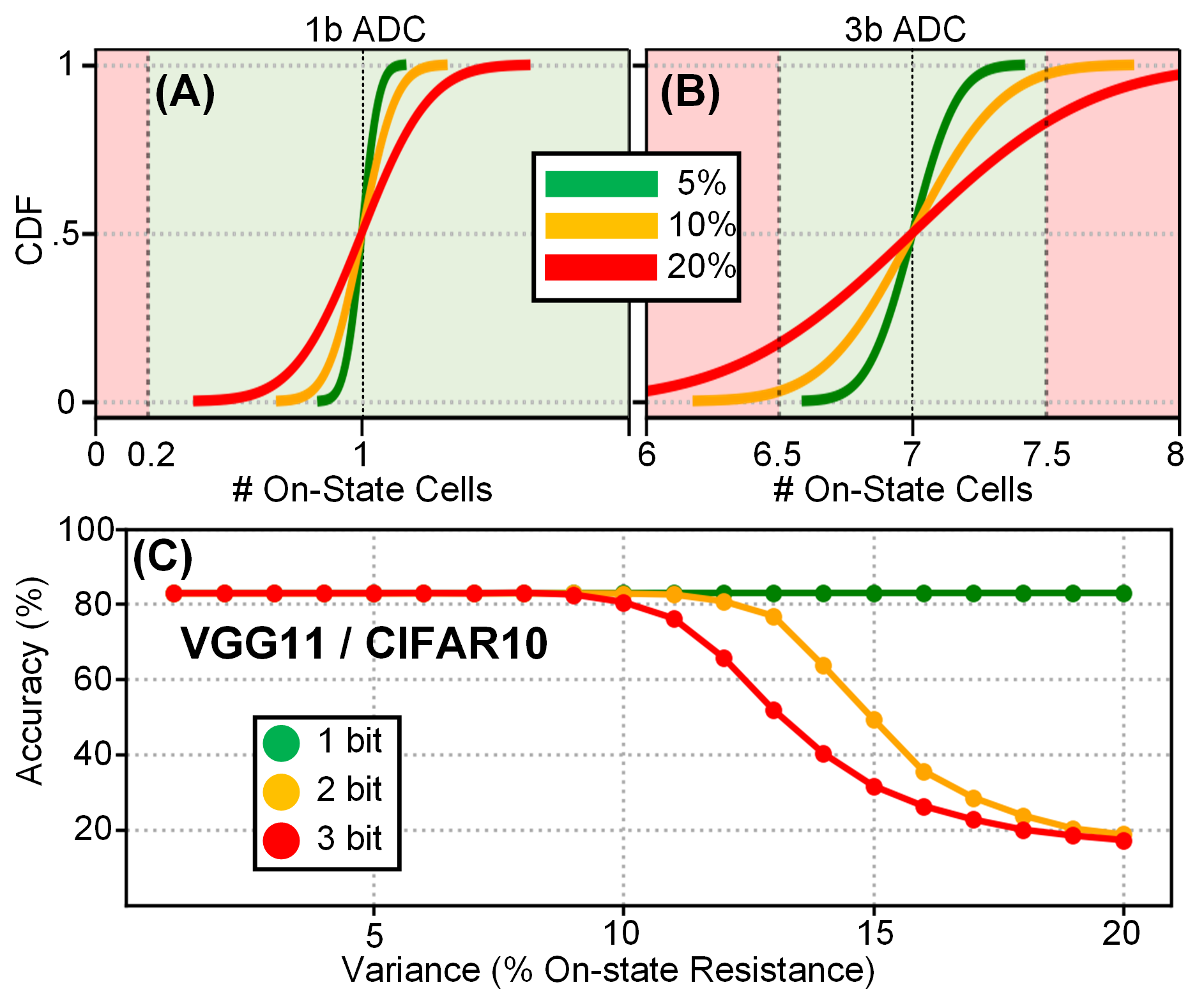}
\vspace{-0.25cm}
\caption{
Impact of ADC precision and variance on CIM. 
(A) Expected distribution when reading 1 on-state device with a 1b ADC. (B) Expected distribution when reading 7 on-state devices with a 3b ADC.
(C) Classification accuracy versus cell-to-cell variance using 1b, 2b, and 3b ADCs. }
\label{fig:var}
\end{figure}

CIM seeks to read and accumulate several states on a bitline at once, and therefore a key obstacle to enabling CIM is cell-to-cell variance. 
Traditional memory cells are read only a single cell on the bitline at once using differential sensing.
Thus deviations in the `0' and `1' states are negligible so long as the ratio between the states is large. 
Unfortunately, because CIM reads several 1 states at the same time, the accumulated variance can easily trigger a $\pm$1 error. 
We illustrate this idea in Figure \ref{fig:var}A and \ref{fig:var}B, where we show 20\% variance is easily tolerated for reading a single cell, but when reading 7 cells at the same time we get a $\pm$1 error frequently. 

Given that CIM is far more prone to errors than traditional single cell read operations, it is important to consider the potential impact of using erroneous computation. 
Interestingly, in many data intensive applications, particularly in neural networks, limited amounts of variance can be tolerated. 
However, it is important to ensure that variance is controlled and does not degrade application accuracy.
To better understand this relationship, we show the impact of variance on classification accuracy in Figure \ref{fig:var}C. 
Using the simulator we discuss in Section \ref{section:results}, we sweep variance and plot classification accuracy in \ref{fig:var}C.
As a example, we implement VGG11 on CIFAR10 with 128$\times$128 arrays.
We observe that as variance increases, absolute error increases linearly. 
As a result, classification accuracy of VGG11 on CIFAR10 falls dramatically after 10\% variance. 

\section{Counting Cards} \label{section:cards}




So far we have discussed cell-to-cell variance and its impact on CIM applications.
%
%
The simplest way to control accumulated variance and error is to reduce the number of word lines enabled at one time. 
Of course, this will unfortunately compromise the performance and energy efficiency of using CIM. 
However, we have little choice in this trade-off because target applications require certain accuracy to be useful. 
%
%
Therefore, our objective function becomes achieving a target accuracy while maximizing performance. 
In this way, the variance of the device should ultimately dictate the precision of the ADC used. 

\begin{figure}
\centering
\includegraphics[width=0.48\textwidth]{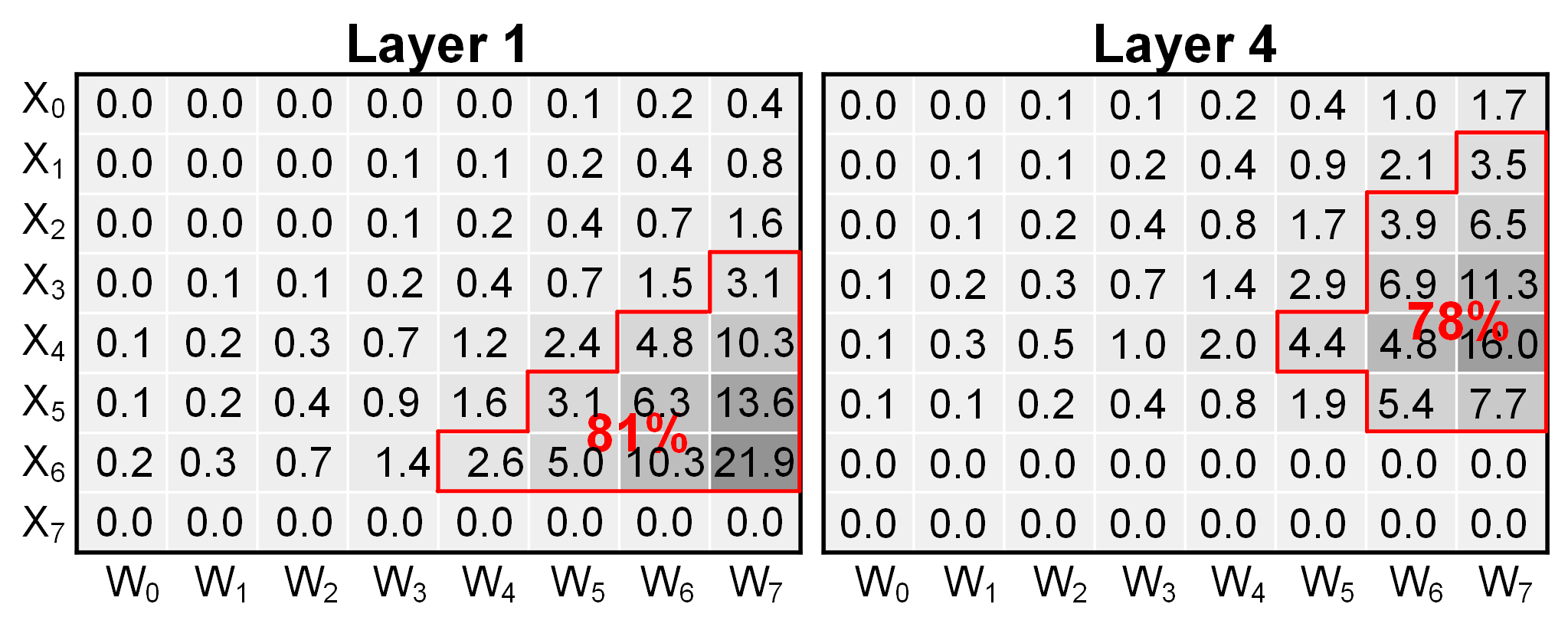}
\vspace{-0.25cm}
\caption{
Each sub-operation's contribution to total error for layers 1 and 4 of VGG11 on CIFAR10.
Note that $X_7$ yields 0 error since the input data is only positive.}
\label{fig:error_table}
\end{figure}

%
CIM arrays perform all sub-operations in binary, and thus all sub-operations are subjected to the same variance and error. 
However, each partial operation has a different magnitude it is multiplied (shifted) with after the binary operation.
Therefore, the majority of error from CIM operations comes from high magnitude partial operations.
To illustrate this idea, we show the error breakdown across partial operations that occurs when performing a 8b matrix multiplication in Figure \ref{fig:error_table}.
Each bin contains the percentage of total error that can be attributed to the sub-operation. 
It is clear that the majority of error comes from high precision values (i.e. 6b$\times$7b), while low precision values contribute near-zero error to the final output. 
If our sub-arrays allow, we can operate these partial operations at different speed (number of wordlines) as a function of their variance and magnitude. 
This idea is the basis of our work, and in the following sections we discuss how to compute expected error, optimize operating speed, and lastly required hardware for operating at different speeds. 



\subsection{Computing Expected Error}



To compute the expected error while reading the array at a certain number of rows per ADC read, we require both application statistics and device level statistics. 
We can take the distribution of LRS and HRS values for a given device and determine the variance measured as the standard deviation of all LRS states. 
To compute the expected error for a matrix multiplication, we can compute the expected error at each individual sub-operation. 
After this, we sum the errors together scaling each one by its corresponding magnitude and dividing by the quantization value associated with the quantized neural network.

For each sub-operation, we read several rows at once, accumulating the results on the bitline before reading out the value from an ADC. 
The more LRS cells that are read, the higher the total variance will be.
This is because LRS is typically significantly larger than HRS ($>$10$\times$), and thus LRS variations have much more impact.
Assuming independent and identically distributed LRS cell resistance, the total variance from reading $N$ LRS cells will be $(\sigma / \mu) \cdot \sqrt{N}$, where $\sigma$ and $\mu$ are the mean and standard deviation of the LRS. 
The higher this total variance, the more likely it is that the value will be read incorrectly by the ADC. 
Assuming a normal distribution, the probability of observing an output,  $\hat{N}$, given a true number of LRS cells, $N$, can be computed using the normal CDF ($\Phi$) function.
This is demonstrated in Equation \ref{adc_error}.

\begin{equation}
\label{adc_error}
P_v(\hat{N} | N) = 
\Phi\left(\frac{\hat{N}-N+0.5}{\sigma\sqrt{N}}\right) - \Phi\left(\frac{\hat{N}-N-0.5}{\sigma\sqrt{N}}\right)
\end{equation}

\begin{figure}
\centering
\includegraphics[width=0.45\textwidth]{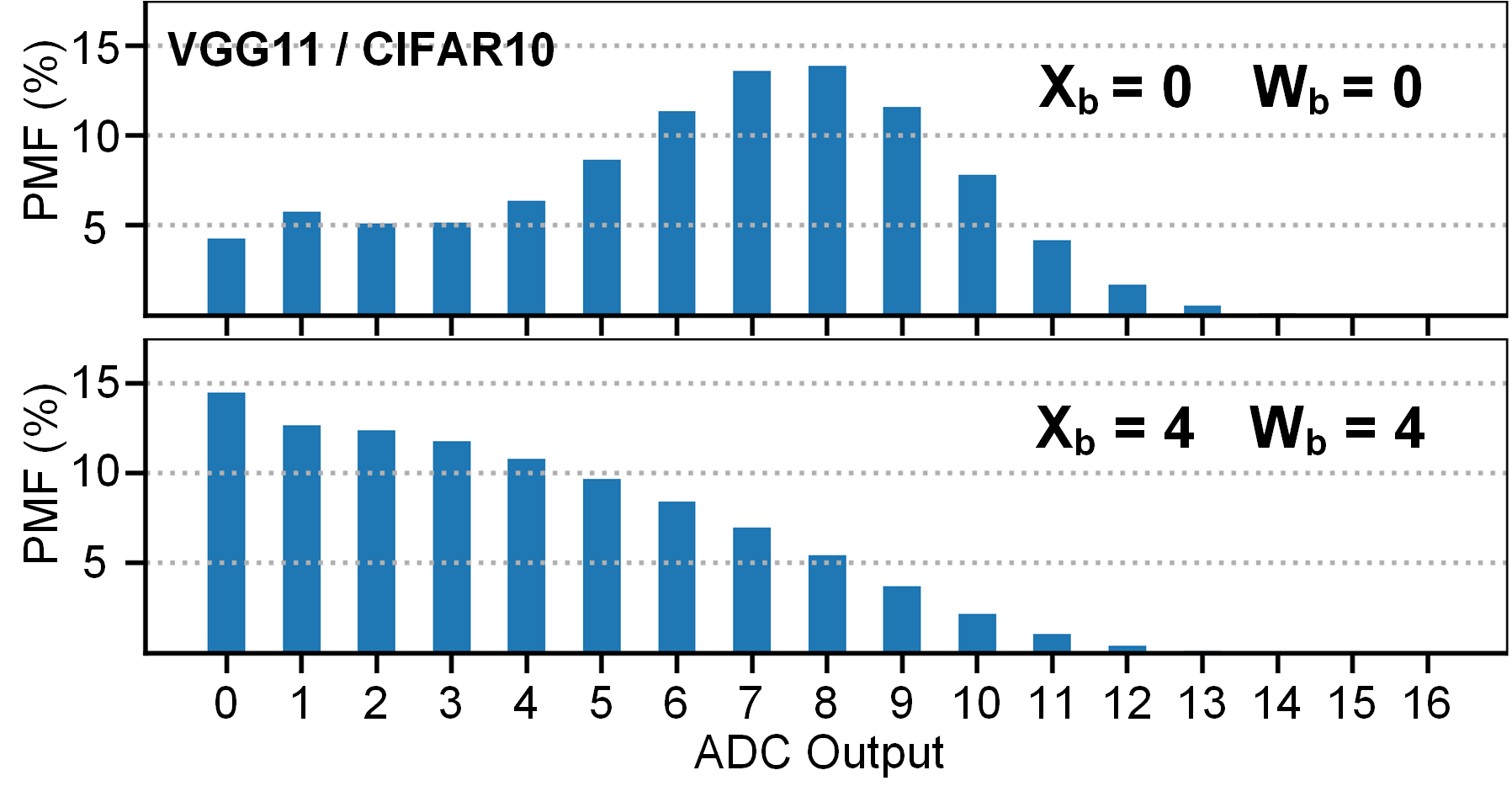}
\vspace{-0.25cm}
\caption{Probability mass function (PMF) for layer 1 of VGG11 on CIFAR10.}
\label{fig:pmf}
\end{figure}

When performing the many ADC reads for a matrix multiplication (or convolution), we will observe a distribution of read values ($N$) that can change as a function of the workload.
Depending on the number of rows enabled, the magnitude of the x and w bits, and the weight vector under consideration, we can expect this distribution to take different forms motivating separate distributions for each condition. 
We are interested in the PMF for $N$, the single-ADC-read true result, so that we can properly resolve the dependence on $N$ in Equation \ref{adc_error}. To acquire this PMF ($P_{xw}(N)$) we can profile the application in simulation, simply counting the number of times each output value occurs. 
As an example, we profile layer 1 of VGG11 on CIFAR10.
We visualize this in Figure \ref{fig:pmf}, showing different distributions for the different sub-operations in the 8b matrix multiplication when reading 16 wordlines at the same time. 
To show the motivation for separate PMFs, we show the product of $X_0 \cdot W_0$ and $X_4 \cdot W_4$.

After acquiring the probability distributions, we can approximate the expected absolute error for a single ADC read by summing the product of the probability of each true and observed readout and the associated error, across all possible true and observed readouts (Equation \ref{exp_error}). The former probability is the product of $P_v(\hat{N} | N)$ and $P_{xw}(N)$ and, naturally, the error associated with the outcome is the absolute difference of the true and observed outputs.

\begin{equation}
\label{exp_error}
E(\epsilon \;|\; x, w) = \sum_N^{N_{WL}} \sum_{\hat{N}}^{N_{WL}} P_v(\hat{N}|N) \cdot P_{xw}(N) \cdot \lvert\hat{N} - N\rvert
\end{equation}

With the expected error for an ADC read in place, we can begin to approximate more abstract operations in the matrix multiplication. 
A matrix multiplication can contain columns larger than $N_{WL}$, so we need to break down each column into several cycles. 
If in a single column we read $N_{tot}$ total cells, we can expect the error to scale sub-linearly with $N_{tot} / N_{WL}$ since in general the errors from single ADC reads sum incoherently.
If we assume linear scaling as an upper bound, we can approximate the error for a column of the matrix multiplication as $(N_{tot} / N_{WL}) \cdot E(\epsilon \;|\; x, w)$.

To approximate the error for matrix multiplication, we must approximate the error for all 64 sub-operations. 
Hence, we can, in the same way, approximate the error of the matrix multiplication as the sum of 64 columns. 
For each X-W pair, we use a unique PMF since the distributions of expected ADC outputs differ between pairs. 
We also multiply the expected error for a single column by the magnitude of the X-W pair. 
Our final upper-bound error approximation for a matrix multiplication is shown in Equation \ref{vmm_error}. 

\begin{equation}
\label{vmm_error}
E_{VMM} = \sum_x^{X} \sum_w^{W} 2^x 2^w \cdot \frac{N_{tot}}{N_{WL}} \cdot E(\epsilon | x, w) 
\end{equation}

\noindent
That is, Equation \ref{vmm_error} approximates the mean absolute error (MAE) for vector-matrix multiplication (VMM). 
Compactly, if we let the true result of VMM be: $y = W x$, and the CIM result be: $\hat{y} = \hat{W} x$, then $E_{VMM}$ computes $\sum \lvert \hat{y} - y \rvert$.






\subsection{ Optimizing Operation Speed }

After establishing a methodology for computing the expected error for a matrix multiplication, we can now formulate an optimization problem. 
In our optimization problem, we consider the relationship between error and performance and we attempt to optimize the performance of our array for a given error threshold. 
This error threshold will be the MAE for VMM.
Examples of MAE thresholds and their relationship to application accuracy are given in Section \ref{section:results}. 

Since each sub-operation will yield different error rates, we choose to operate each one at a different number of wordlines per cycle. 
Thus, the solution to our optimization problem will be a lookup table (LUT) containing 64 ($8\times8$) values indicating the number of wordlines per cycle to be performed by the corresponding sub-operation. 
Due to the immense difference in magnitude between the low and high magnitude operations, a better performance can be achieved by enabling more wordlines than they have comparators to sense. 
Doing so means we will incur quantization error because it is not possible to sense more states than we have comparators in our ADC. 
While this will naturally incur error, we can compute this expected error the same way we have done so far and enable better performance while satisfying our threshold. 

\begin{figure}
\centering
\includegraphics[width=0.45\textwidth]{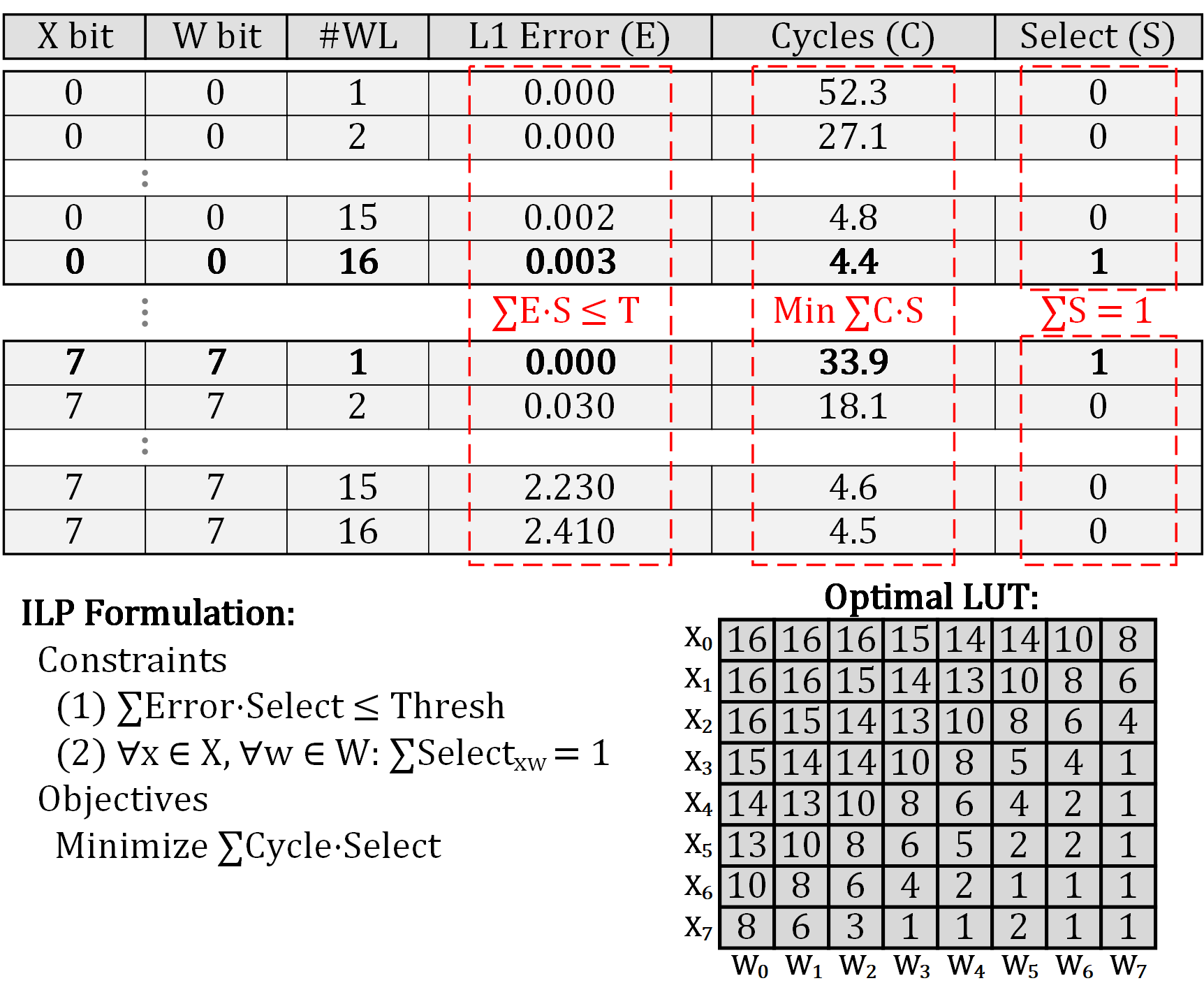}
\vspace{-0.25cm}
\caption{
Example LUT optimization problem formulation. }
\label{fig:ilp}
\end{figure}

To find the optimal LUT, we compute the expected error and performance for all sub-operations operating at various numbers of wordlines per cycle. 
We then compile these results into a table.
To illustrate this, we provide an example table in Figure \ref{fig:ilp}. 
In this example, we perform 8b matrix multiplication and we read up to 16 wordlines at a time, thus our table will be $8\times8\times16$. 
For each sub-operation, we must select between 1 wordline and 16 wordlines in a single cycle.
This is visualized in the far right column of Figure \ref{fig:ilp}, where one of the 16 conditions will be chosen. 
The objective of this problem is to satisfy some error threshold, while maximizing performance. 

Upon formulating this problem we note that it is, by definition, a variant of the knapsack problem called multi-class knapsack. 
In this instance, the weight of each sub-operation is the error, the knapsack's capacity is the error threshold, and the value is the delay required for a partial operation.
In this case, the objective function is minimization since we seek to minimize delay.
The result is the number of wordlines to be enabled for each partial operation. 
The constraints are two fold. 
First, the sum of the error of all the selected sub-operations must be less than or equal to our error threshold.
Second, exactly 1 speed must be chosen for each of the 64 sub-operations, hence multi-class knapsack. 
This can be solved with a greedy heuristic or a common optimization technique like branch and bound or integer linear programming (ILP). 
We chose to use ILP since it required the simplest implementation and executes very fast ($<$ 1s). 

\subsection{ Hardware Support }

To set the speed of our arrays for each sub-operation, we require additional hardware. 
Each array must be controlled by a LUT specifying the number of wordlines to be read each cycle. 
These LUTs can be shared with all sub-arrays in our design, however we put a duplicate LUT in each PE (64 sub-arrays) to reduce high fan-out. 
To set the speed we require $\log(N)$ bits per sub-operation, where $N$ is the maximum number of wordlines we will enable in our sub-array.
For example, if we enable up to 16 wordlines, we need 4b per sub-operation.
It should be noted that sub-operations can be grouped together to reduce hardware complexity.
For example, in 8b matrix multiplication, we use 16 groups of 4 sub-operations rather than 64 sub-operations to reduce register footprint by 4. Since we modulate speed only via the number of enabled wordlines, without adjusting the clock period, there is no need for additional clocking hardware overhead.

\section{CIM-based Architecture} \label {section:circuit}

For our experiments we adopted a similar architecture to previous work \cite{shafiee2016isaac}.
Our basic processing element (PE) contains 64 256$\times$256 arrays. 
We choose 64 arrays because it provides each block with sufficient network bandwidth and SRAM capacity, while maintaining good SRAM density and low interconnect overhead. 
In all designs we consider, we use the same 64 array PE and simply increase the count per design. 
Our input data, weights, and activations are all 8 bits.
Each array has 1 3-bit ADC for every 8 columns where a single column is pitch-matched with a comparator. 
In Figure \ref{fig:table1}, we show the relevant specifications of our sub-array and processing element and the assumed energy per operation. 
This includes the WL LUT specific to this work that is shared across all 64 arrays per PE.

\begin{figure}
\centering
\includegraphics[width=0.45\textwidth]{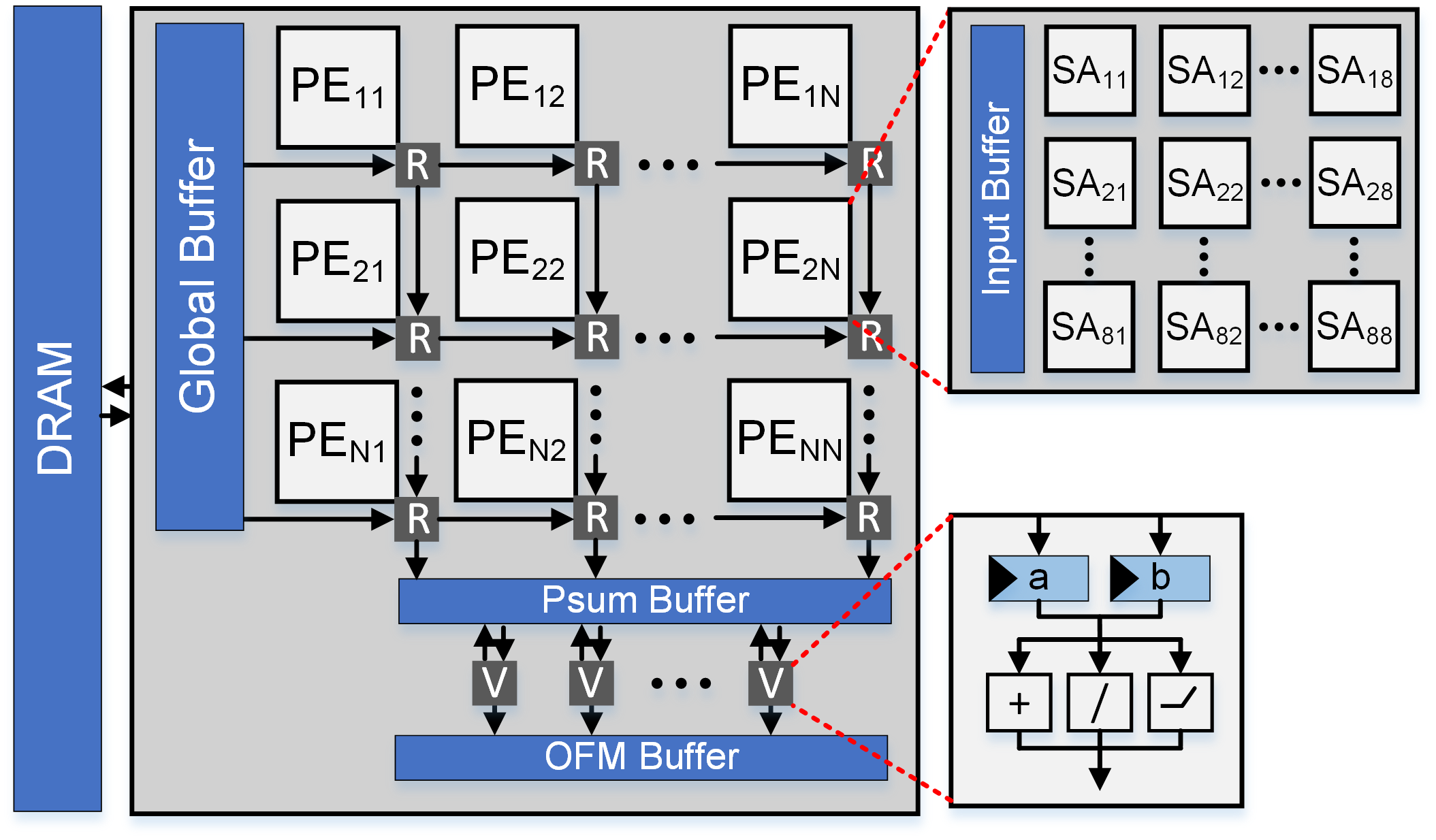}
\vspace{-0.25cm}
\caption{
System architecture with 1 router (R) per PE. 
All input features are routed from the global buffer to PEs. 
All partial sums are routed from PE to vector unit (V), and vector unit to output feature buffer.
}
\label{fig:network}
\end{figure}

The activation inputs to the RRAM sub-arrays are stored in on-chip SRAM, while the input images are read in from external DRAM.  
Matrix multiplication is performed by the PEs, while custom vector units are used to perform vector-wise accumulation, bias addition, quantization, and relu.
We use a $N \times N$ mesh network for communication between PEs, memory, and vector units shown in Figure \ref{fig:network}. 
Because of the high write energy cost of eNVM, we adopt layer pipelining and weight duplication \cite{shafiee2016isaac}. 
To partition the weights and maximize throughput we adopt the allocation policy proposed in \cite{crafton2020breaking}. 


\section{Results} \label {section:results}

\begin{figure}
\centering
\includegraphics[width=0.45\textwidth]{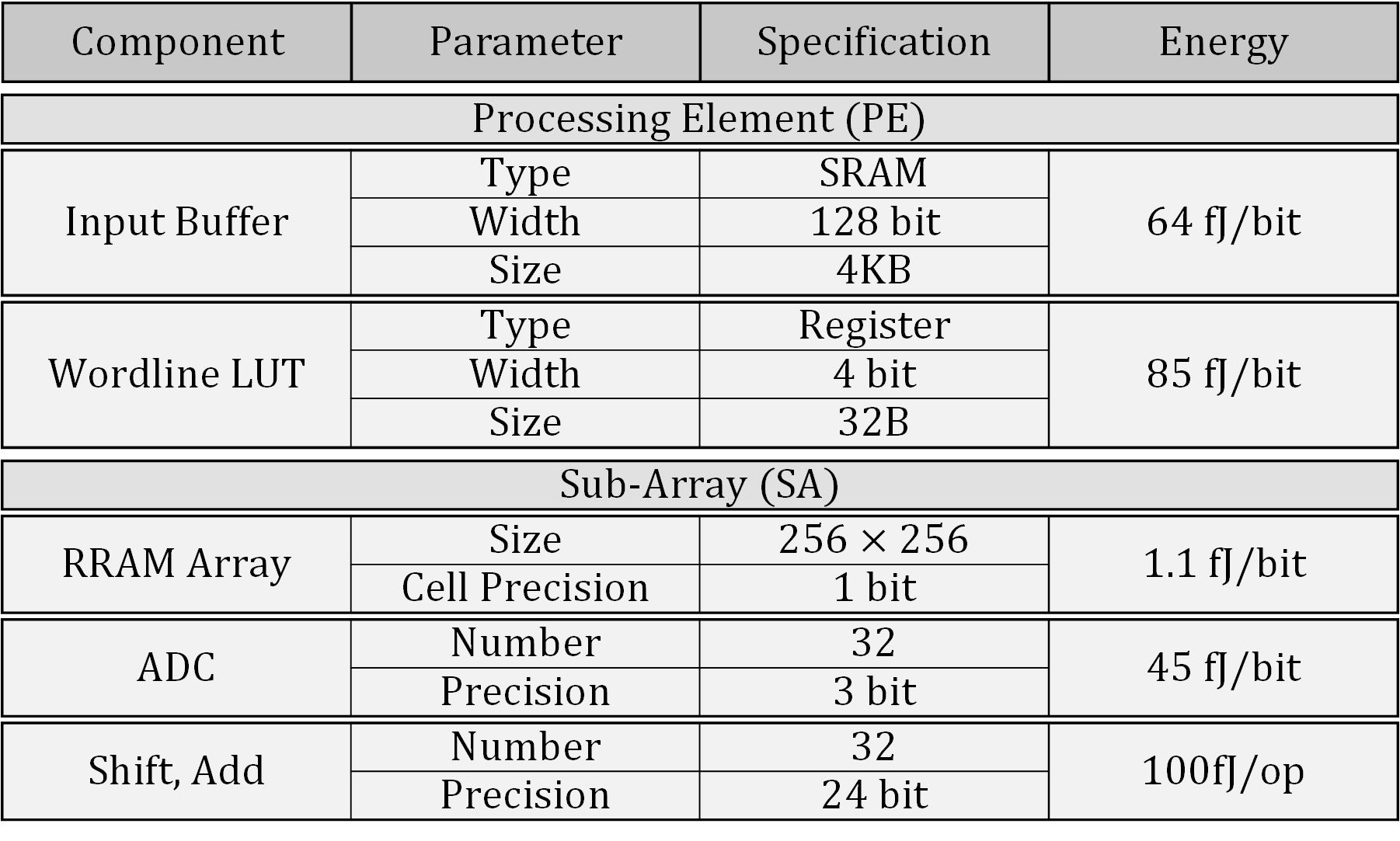}
\vspace{-0.25cm}
\caption{Simulation parameters used for hardware components at both the sub-array and processing element level.}
\label{fig:table1}
\end{figure}

To benchmark our algorithm, \textit{counting cards}, we compare against the two commonly used techniques: baseline and zero-skipping. 
We empirically evaluate performance, power, and accuracy for the three techniques on ImageNet using ResNet18 and CIFAR10 using VGG11. 
Both models were trained from scratch and use 8b weights and activations.
ResNet18 obtains a top-1 accuracy of 68.4\% on ImageNet, and VGG11 obtains a top-1 accuracy of 84.75\% on CIFAR10. 
We run these techniques in a custom simulation framework designed to evaluate performance and power of CIM using standard 32nm CMOS and RRAM models adopted from \cite{chen2017neurosim+}, and displayed in Figure \ref{fig:table1}.
For both benchmarks, we use a 64 PE (4096 array) system based on the parameters in Figure \ref{fig:table1}.

Our simulator performs cycle-accurate implementations of convolutional and fully connected layers. 
It is based in Python, but runs array level operations in C for faster evaluation. 
We model components in the design in object oriented fashion, iterating through all components in all PEs each cycle.
We embed performance counters in our ADC and sub-array objects to track metrics like stalls so we can calculate utilization. 
As input, the simulator takes the network weights, input images, PE level configuration, and chip-level configuration. 
The PE-level configuration includes details like the precision of each ADC and size of the sub-array. 
The chip-level configuration contains the number of PEs and details about array allocation and mapping.
As output, the simulator produces a table with all desired performance counters and all intermediate layer activations that are verified against a TensorFlow implementation for correctness. 

\begin{figure}[t]
\centering
\includegraphics[width=0.45\textwidth]{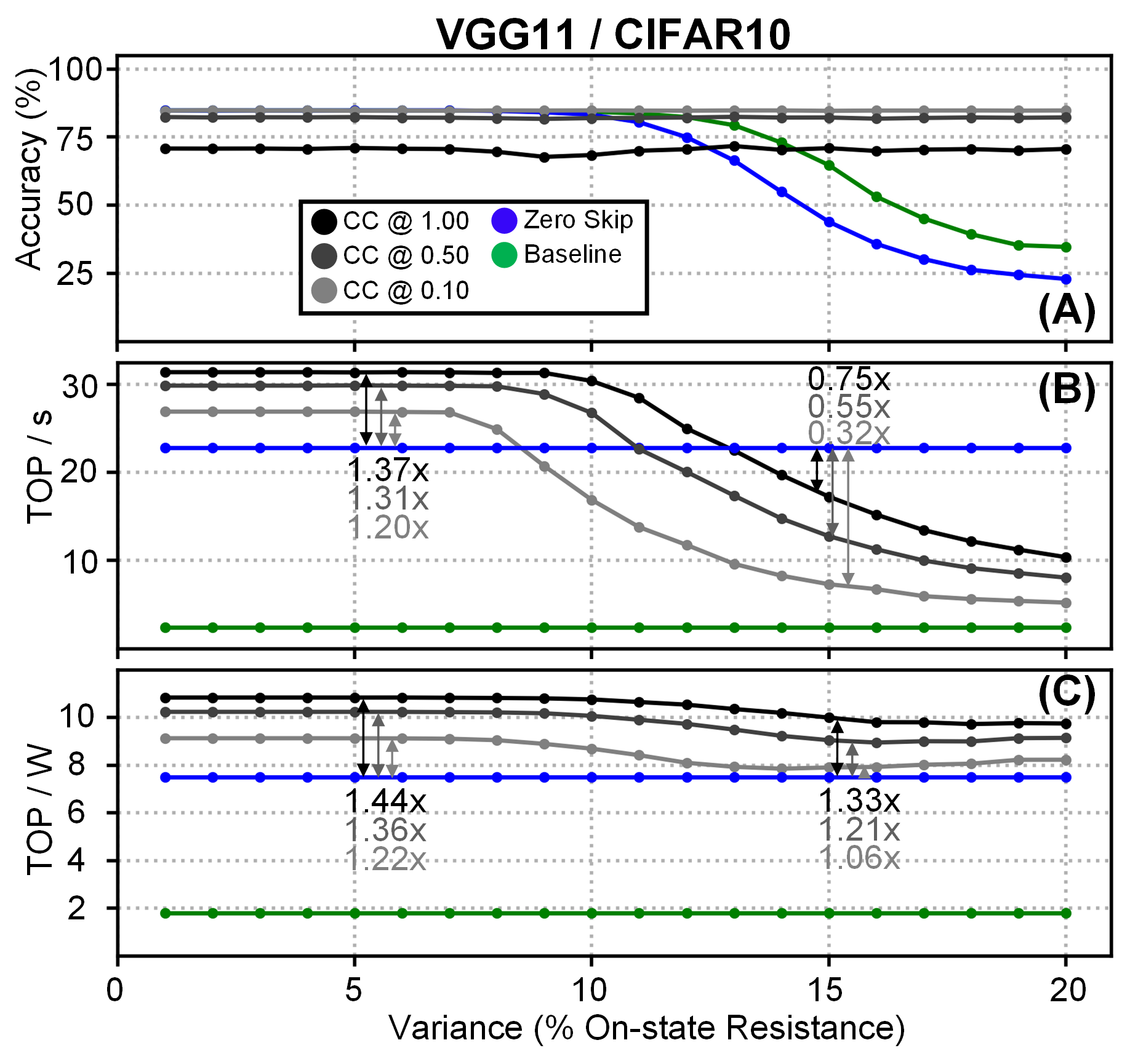}
\vspace{-0.25cm}
\caption{
Classification accuracy, performance, and energy efficiency versus cell-to-cell variance for VGG11 on CIFAR10 assuming 100MHz clock. 
}
\label{fig:cc1}
\end{figure}

\begin{figure}[t]
\centering
\includegraphics[width=0.45\textwidth]{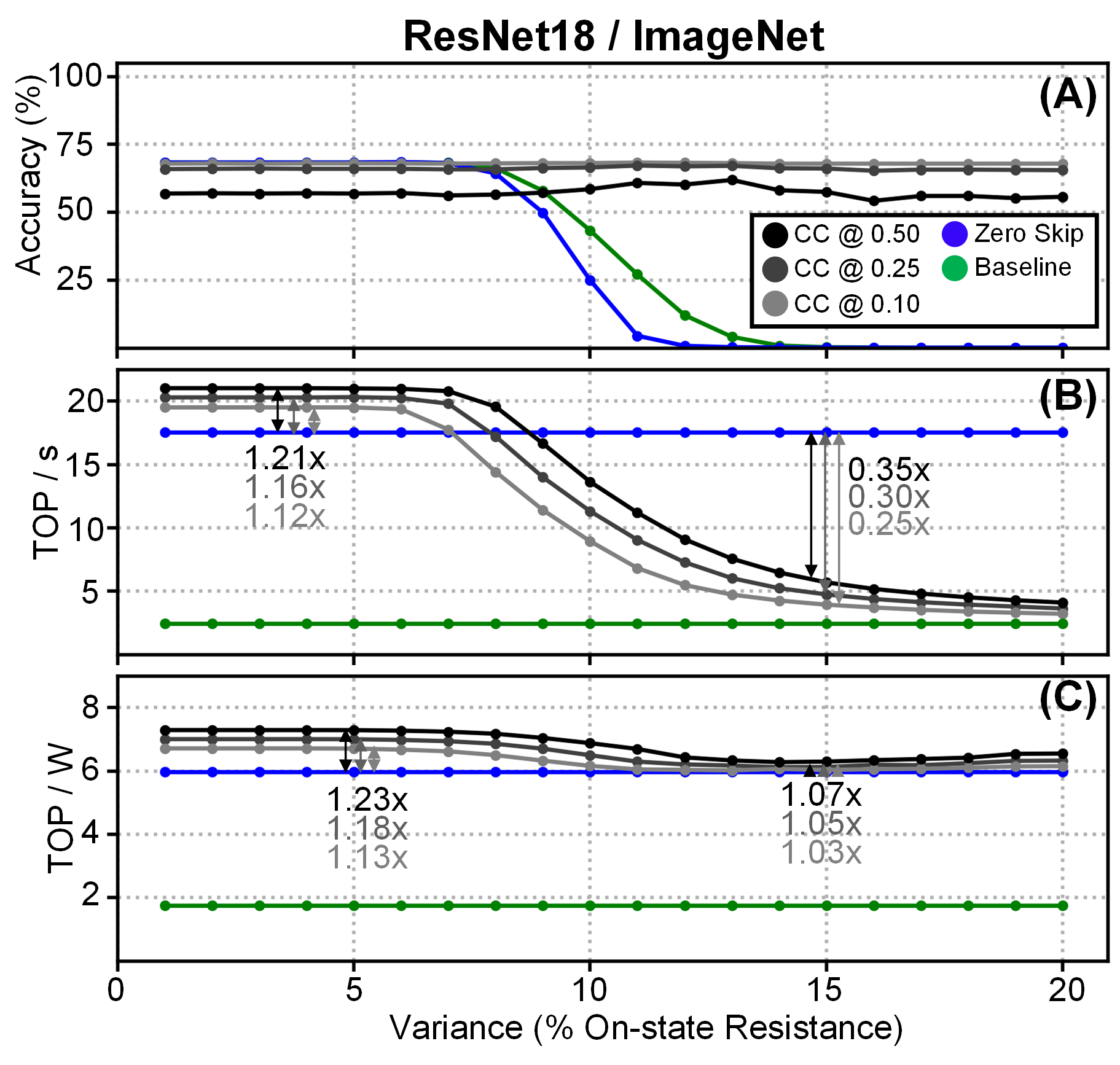}
\vspace{-0.25cm}
\caption{
Classification accuracy, performance, and energy efficiency versus cell-to-cell variance for ResNet18 on ImageNet assuming 100MHz clock. 
}
\label{fig:cc2}
\end{figure}

To show how \textit{counting cards} performs under various conditions, we have evaluated accuracy, performance, and energy efficiency over different device variance and error thresholds. 
In Figures \ref{fig:cc1} and \ref{fig:cc2}, we plot our results for CIFAR10 and ImageNet, respectively.
We show counting cards running at three different error thresholds to show the trade-off between accuracy and error. 
The three different thresholds reveal the sensitivity of the network to variations, which are very difficult to predict mathematically, and thus empirical simulations provide sufficient information to select an error threshold. 

\begin{figure}
\centering
\includegraphics[width=0.48\textwidth]{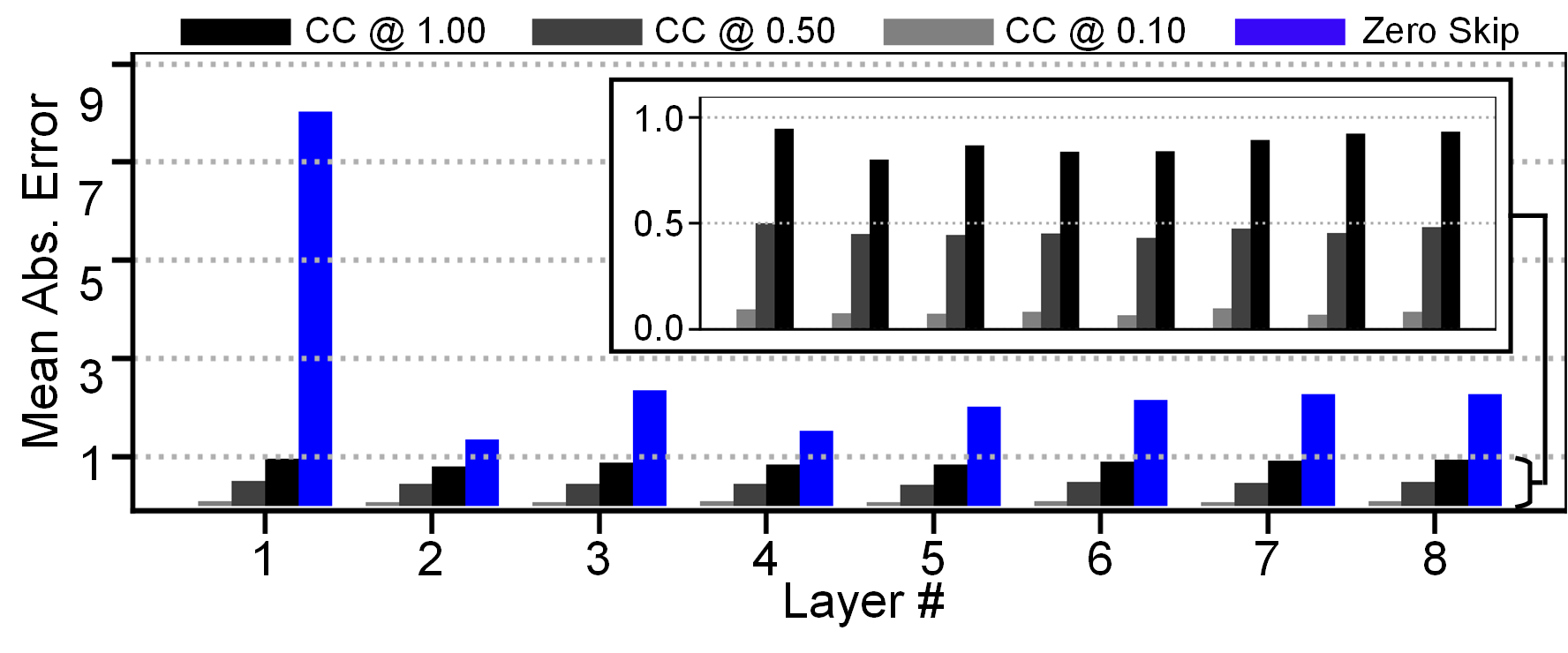}
\vspace{-0.25cm}
\caption{Mean absolute error by layer for VGG11 on CIFAR10.
}
\label{fig:mae}
\end{figure}

We begin at 1\% variance and sweep up to 20\% variance.
At low variance (1\%-5\%), the counting cards methods yield higher performance than the traditional methods, since we operate sub-operations at more wordlines than we have ADCs. 
This results in accuracy degradation for counting cards with error thresholds at 0.5 and 1. 
However, we observe that there is no significant degradation for counting cards with 0.1 error threshold.  
As variance increases, the traditional methods begin to incur error and the accuracy begins to drop. 
For counting cards, error remains fixed, and instead performance degrades to compensate for the higher variance. 
Across all variance configurations, we observe an energy advantage to using counting cards. 
This is because for many partial operations we enable more wordlines than we have ADC states, and thus we perform more work per read. 
In Figure \ref{fig:mae}, we show MAE by layer for 20\% variance on CIFAR10. 
This figure demonstrates how counting cards keeps precise control of MAE at each layer based on the error threshold, while the zero skip exhibits high error across all layers. 

We follow the same procedure for ResNet18 on ImageNet, however we observe that counting cards yields less performance and energy efficiency advantages. 
This is due to a couple reasons. 
First, ResNet18 has more layers, and error at each layer propagates throughout the whole network meaning lower error thresholds at each layer must be used. 
Furthermore, ImageNet is a far more complex dataset and is more sensitive to errors.
This is observed in Figure \ref{fig:cc2}A, where accuracy for zero-skipping and baseline degrades much faster than it did for CIFAR10. 
Second, ResNet18 contains much larger feature maps ($224\times224$) than CIFAR10 ($32\times32$).
These early layers contain higher percentages of `1's in the feature maps and thus more ADC reads per operation are required. 

\section{Conclusion} \label {section:conclusion}



In this paper we demonstrate the efficacy of a new technique, counting cards, to control the impact of variance in CIM accelerators. 
Using weight and activation statistics from the target application as well as cell-to-cell variance, we can compute the optimal performance for a given error threshold.  
Experimental results show counting cards yields a 36\% energy improvement and 31\% performance improvement over existing work, while satisfying a programmable error threshold.


\section{Acknowledgement} \label {acknowledgement}
\noindent
This work was funded by the U.S. Department of Defense’s Multidisciplinary University Research Initiatives (MURI) Program under grant number FOA: N00014-16-R-FO05 
and the Semiconductor Research Corporation under the Center for Brain Inspired Computing (C-BRIC) 
and Qualcomm. 
\small{
\bibliographystyle{ieeetr}
\bibliography{main}
}
\end{document}